\documentclass[twocolumn,prl,showpacs,amsmath,amssymb]{revtex4-1}
\usepackage{graphicx}
\usepackage{dcolumn}
\usepackage{bm}
\usepackage{epsfig}

\begin{document}
\title{Gravity gradient suppression in spaceborne atomic tests of the equivalence principle}
\author{Sheng-wey Chiow, Jason Williams, Nan Yu}
\email{nan.yu@jpl.nasa.gov}
\affiliation{Jet Propulsion Laboratory, California Institute of Technology, Pasadena, CA 91109}
\author{Holger M\"uller}
\email{hm@berkeley.edu}
\affiliation{Department of Physics, University of California, Berkeley, CA 94720}

\date{\today}

\begin{abstract}
The gravity gradient is one of the most serious systematic effects in atomic tests of the equivalence principle (EP). 
While differential acceleration measurements performed with different atomic species under free fall test the validity of EP, minute displacements between the test masses in a gravity gradient produces a false EP-violating signal that limits the precision of the test.
We show that gravity inversion and modulation using a gimbal mount can suppress the systematics due to gravity gradients caused by both moving and stationary parts of the instrument as well as the environment, strongly reducing the need to overlap two species.
\end{abstract}

\maketitle

The equivalence principle (EP) is one of the pillars of the general theory of relativity, and has become a touchstone for quantum gravity, dark matter, and dark energy~\cite{QTEST}.
Precision tests of the EP are powerful tools to probe Planck-scale physics at low energy, and physics beyond the standard model~\cite{damour}, including the low-energy limit of a possible theory of quantum gravity~\cite{kostelecky2004gravity,kostelecky2011matter,bailey2015short} and searches for dark matter~\cite{PhysRevD.93.075029} and dark energy~\cite{hamilton2015atom}.
The EP was tested at a precision of $\sim 10^{-10}$ by dropping bulk masses~\cite{carusotto1992test}, and at $\sim 10^{-13}$ by using a torsion balance~\cite{wagner2012torsion}.
Quantum tests of the Universality of Free Fall (UFF) by comparing accelerations of different atomic species with atom interferometry (AI) have been proposed~\cite{Hogan2008light} and conducted~\cite{Zhou_2015,BouyerEP} at $\sim10^{-9}$ by several research groups, as summarized in Ref~\cite{QTEST}. 
The extended observation times and improved control of systematic effects in spaceborne AI promise orders of magnitude improvements in sensitivity~\cite{QTEST}.
In such equivalence principle tests the gravity gradient is the most serious systematic effect, where species-dependent acceleration arises due to imperfect spatial overlap of test masses~\cite{QTEST,uscope,touboul2012microscope,Hogan2008light}.
Various measures are necessary to control this systematic, e.g., extensive in-flight calibrations~\cite{uscope,touboul2012microscope} or environmental control~\cite{Hogan2008light}.
However, the effectiveness of these mitigations is not all obvious.
There is even a recent claim asserting the existence of a fundamental measurement limit in the presence of gravity gradients, as a result preventing a high precision EP measurement~\cite{Nobili}.

\begin{figure*}[t]
\centering
\includegraphics[width=\textwidth]{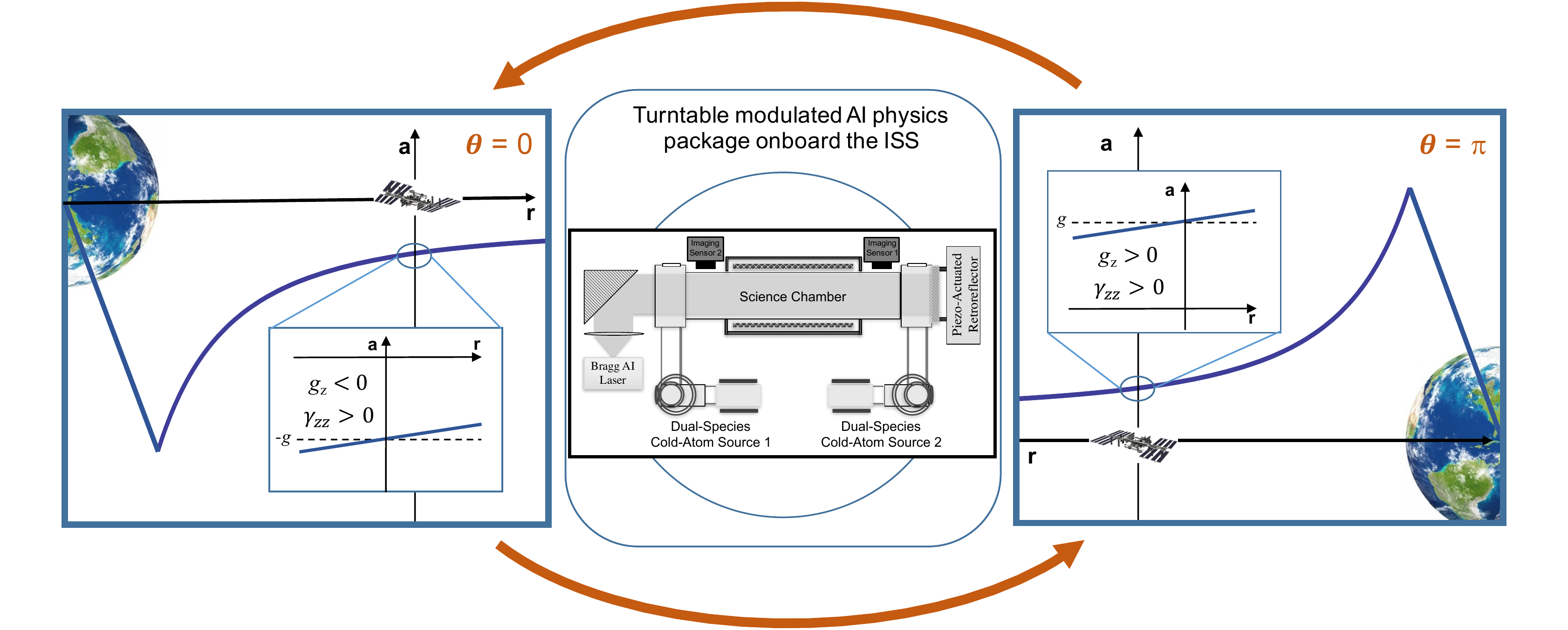}
\caption{Illustration of the gravity modulation scheme in the context of the QTEST concept~\cite{QTEST}. 
The center figure is the atom interferometer EP experimental setup on a turntable which is modulated between 0 and $\pi$ along the nadir direction ($r$). 
In the reference frame of the turntable (experiment), the projection of the Earth's gravitational acceleration along $z$ ($g_z$), and therefore the EP signal ($\eta\ k_{\rm{eff}}g  T^2$), changes sign as the turntable is modulated from $\theta = 0$ (left) to $\theta = \pi$ (right). 
On the other hand, the gravity gradient from the Earth, $\gamma_{zz}$, remains the same under the modulation (by definition, the gravity and the gravity gradient of the instrument remain constant in the experiment reference frame). 
Therefore, the gravity gradient effects can be effectively suppressed in differential measurements. 
The insets zoom to the apparatus size, emphasizing that the modulation changes the relative sign between the gravitational acceleration ($g_z$) and the gravity gradient ($\gamma_{zz}$) seen by the experiment.}\label{fig1}
\end{figure*}

In this letter, we show that inversion of gravity by rotating the AI apparatus on a gimbal can suppress the sensitivity to gravity gradients caused by both the rotating parts of the instrument itself and the environment, strongly reducing the requirements on the overlap of the two species (this concept has been employed for bulk test masses~\cite{reasenberg2014new,uscope,touboul2012microscope}). 
An example of this scheme, utilizing the International Space Station (ISS) for dual-species AI experiments in microgravity~\cite{QTEST}, is illustrated in the instrument frame of reference in Fig.~\ref{fig1}.
The suppression is based on the fact that inversion of the instrument leaves the gravity gradient tensor unchanged in its principal-axis frame, while the gravitational acceleration changes sign and can thus be measured independently, if the apparatus (in particular spatial mismatch between the atomic samples) is otherwise unaffected by the inversion.
This is the case in microgravity with good magnetic shielding.
The gravity modulation thus greatly suppresses systematics and their variations.

The significance of the gravity gradient for AI-based EP tests is briefly summarized as follows.
The phase $\phi$ of an AI in the Mach-Zehnder configuration is (to the leading orders):
\begin{equation}\label{eq1}
\phi=\vec{k}_{\rm{eff}}\cdot \vec{g}\ T^2+\vec{k}_{\rm{eff}}\cdot \overset{\text{\tiny$\bm\leftrightarrow$}}{\gamma} \cdot\left(\vec{z}_i+\vec{v}_i T\right)T^2+\cdots,
\end{equation} where $\vec{k}_{\rm{eff}}$ is the effective wavevector of the AI laser pulses (direction fixed to the apparatus), $\vec{g},\overset{\text{\tiny$\bm\leftrightarrow$}}{\gamma}$ are the first and the second order derivatives of the gravitational potential (also commonly recognized as the gravitational acceleration vector and the gravity gradient tensor), $T$ is the time between AI laser pulses, and $\vec{z}_i,\vec{v}_i$ are the initial position and velocity of the atom~\cite{QTEST,Hogan2008light}.
In practice, the effective location and velocity of an extended atomic sample are hard to control.
Typically, $\vec{k}_{\rm{eff}}$ is chosen parallel to $\vec{g}$ to maximize the sensitivity~\footnote{In microgravity, where $\vec{g}$ is balanced by the centrifugal force in the moving frame, $\vec{k}_{\rm{eff}}$ is chosen to point vertically so that it is parallel to $\vec{g}$ in an inertial frame.}. 
In general, $\overset{\text{\tiny$\bm\leftrightarrow$}}{\gamma}=\overset{\text{\tiny$\bm\leftrightarrow$}}{\gamma}_\oplus+\overset{\text{\tiny$\bm\leftrightarrow$}}{\gamma}_{\textrm{S}}$ consists of the Earth's gravity gradient $\overset{\text{\tiny$\bm\leftrightarrow$}}{\gamma}_\oplus$ and the gravity gradient of the instrument $\overset{\text{\tiny$\bm\leftrightarrow$}}{\gamma}_{\textrm{S}}$, the self gravity gradient (SGG).

The E\"otv\"os parameter $\eta$ is defined as
\begin{equation}
\eta=2\frac{g_A-g_B}{g_A+g_B},
\end{equation} where $g_A$ ($g_B$) is the gravitational acceleration measured with atomic species $A$ ($B$).
Due to the gravity gradient, a direct comparison of AI phases of different species will have nonzero signal even if $\eta=0$ and $g_A=g_B$:
\begin{equation}\label{eq3}
\delta \phi_{\theta=0}=\eta\ \vec{k}_{\rm{eff}}\cdot \vec{g} \ T^2+ \vec{k}_{\rm{eff}}\cdot\overset{\text{\tiny$\bm\leftrightarrow$}}{\gamma}\cdot\left(\delta \vec{z}_i+\delta \vec{v}_i T\right)T^2+\cdots,
\end{equation} where $\delta \vec{z}_i$ ($\delta \vec{v}_i$) is the initial position (velocity) difference of the two species.
With Earth's nominal gravity gradient $\gamma_\oplus\simeq 3000$ nm/s$^2$ per meter in the radial direction, establishing a bound of $10^{-15}$ on $\eta$ would require overlapping of the two atomic clouds with a high accuracy of $\delta z_i = 10^{-15} g/\gamma_\oplus\simeq 3$ nm, which is technically difficult to achieve or even measure due to the spatial extension of atomic clouds.
A similar requirement on initial velocity mismatch also exists.

For clarity, let's consider the one-dimensional case that $\vec{k}_{\rm{eff}}$ is aligned vertically to $\vec{g}$ and therefore parallel to the major principal axis of $\overset{\text{\tiny$\bm\leftrightarrow$}}{\gamma}_\oplus$, and the gimbal rotation is ideally 180$^\circ$ about a second principal axis of $\overset{\text{\tiny$\bm\leftrightarrow$}}{\gamma}_\oplus$.
In this ideal case, $\delta \phi_{\theta=0}=\eta\ k_{\rm{eff}}g \ T^2+ k_{\rm{eff}}\gamma\left(z_i+v_i T\right)T^2+\cdots$.
As seen in the apparatus frame, $g\rightarrow -g$ when the gimbal is turned, while $k_{\rm{eff}} ,\gamma_{\textrm{S}},z_i,v_i$ remain the same.
The gravity gradient $\gamma_\oplus$, too, remains the same:
\begin{eqnarray}\label{eq6}
\gamma_\oplus' &\equiv &\left.\frac{\partial^2 }{\partial z^2}\Phi_\oplus\left(L_t-z\right)\right|_{z=L_t}\\ \nonumber
&=&\left.\frac{\partial^2 }{\partial (L_t-z)^2}\Phi_\oplus\left(L_t-z\right)\right|_{z=L_t}\\ \nonumber
&=&\left.\frac{\partial^2 }{\partial z^2}\Phi_\oplus\left(z\right)\right|_{z=L_t}\\ \nonumber
&=&\gamma_\oplus,
\end{eqnarray}where $\Phi_\oplus$ is the Earth's gravitational potential, and $L_t$ is the turning point of the gimbal and the origin of the coordinate system.
The phase difference of two AIs measured at the flipped orientation is
\begin{equation}\label{eq7}
\delta \phi_{\theta=\pi}=-\eta\ k_{\rm{eff}}g\ T^2+k_{\rm{eff}}\gamma\left(\delta z_i+\delta v_i T\right)T^2+\cdots.
\end{equation}
The difference $\delta \phi_{\theta=0}-\delta\phi_{\theta=\pi}=2\eta\ k_{\rm{eff}}g\ T^2$ is immune to $\gamma$.
The result is still valid when considering only part of the instrument is rotating on the gimbal: the potential of the non-rotating part of the instrument, including the spacecraft and housing, can be combined with the Earth's potential, thus the gravity gradient remains the same under flipping.

Extending the analysis to three dimensions and considering a finite inversion imperfection angle $\delta \theta \ll 1$, the phase difference becomes:
\begin{eqnarray}\label{eq8}
\delta \phi_{\theta=0}-\delta\phi_{\theta=\pi} & \simeq & 2\eta\ k_{\rm{eff}}g\ T^2 \\ \nonumber
&-& 2k_{\rm{eff}}\ {\gamma_\oplus}_\perp\left(\delta z_i+\delta v_i T\right)_\perp \delta \theta \ T^2\\ \nonumber
&+ &  2k_{\rm{eff}}\  {\gamma_\oplus}_\|\left(\delta z_i+\delta v_i T\right)_\| \delta \theta^2 \ T^2\\ \nonumber
&+& \cdots, 
\end{eqnarray} where the subscript $_\|$ indicates the component in the $\vec{k}_{\rm{eff}}$ direction, and the subscript $_\perp$ indicates the component in the direction orthogonal to both $\vec{k}_{\rm{eff}}$ and the turning axis.
Even taking experimental imperfections into account, a suppression factor of 10000 can still be obtained with $\delta \theta=0.1$~mrad inversion angle error.

\begin{table*}
\begin{tabular}{c| c}
Phase term & Relative magnitude \\
\hline
\hline
$ \frac{1}{2} k_{\text{eff}}  (T_{\text{zz}}+t_\text{zz}) ((\delta z_1+\delta z_2)+(\delta v_{z1}+\delta v_{z2}) T)T^2 \cos\theta \cos\psi$ & $7.7\times10^{-12}{}^{\ C}$ \\
$ -k_{\text{eff}} (\delta v_{x1}+\delta v_{x2}) \delta \Omega_y T^2 \cos\theta \cos\psi $&$ 2.6\times 10^{-13}{}^{\ *} $\\
$ -\frac{5}{6} k_{\text{eff}} (T_{\text{xx}}+t_\text{xx})(\delta v_{x1}+\delta v_{x2}) \Omega_y T^4  \cos\theta \cos\psi $&$1.0\times 10^{-13}{}^{\ *} $\\
$ \frac{7}{12} k_{\text{eff}} (T_{\text{zz}}+t_\text{zz}) (\delta v_{x1}+\delta v_{x2}) \Omega_y T^4 \cos\theta \cos\psi$&$9.2\times 10^{-14} {}^{\ *}$\\
$ \frac{1}{2} k_{\text{eff}}( (T_\text{xz}+t_\text{xz})(\delta v_{x1}+\delta v_{x2})+(T_\text{yz}+t_\text{yz})(\delta v_{y1}+\delta v_{y2}) )  T^3 \cos\theta \cos\psi$&$8.6\times 10^{-14} {}^{\ B}$\\
$ -4 k_{\text{eff}} (\delta v_{y1}+\delta v_{y2}) \delta \Omega_z \Omega_y T^3 \cos\theta \cos\psi$&$1.2\times 10^{-14}{}^{\ A} $\\
$ -\frac{3}{4} k_{\text{eff}}(T_{\text{xx}}+t_\text{xx}) \Omega_y (\delta x_1+\delta x_2) T^3  \cos\theta \cos\psi$&$9.4\times 10^{-15} {}^{\ *}$\\
$ \frac{1}{2} k_{\text{eff}} ( (T_\text{xz}+t_\text{xz})(\delta x_1+\delta x_2)+(T_\text{yz}+t_\text{yz})(\delta y_1+\delta y_2) )T^2  \cos\theta \cos\psi$&$8.6\times 10^{-15}{}^{\ B} $\\
$ -\frac{1}{2} k_{\text{eff}} (T_{\text{xx}}+t_\text{xx})(\delta v_{x1}+\delta v_{x2}) T^3  \sin\theta$&$5.5\times 10^{-15}  {}^{\ F}$\\$
 \frac{1}{2} k_{\text{eff}}  (T_{\text{yy}}+t_\text{yy})(\delta v_{y1}+\delta v_{y2}) T^3 \cos\theta \sin\psi$&$5.5\times 10^{-15} {}^{\ *}$\\$
 -\frac{53}{48} k_{\text{eff}} (T_{\text{zz}}+t_\text{zz}) (\delta v_{z1}+\delta v_{z2}) \Omega_y^2 T^5 \cos\theta \cos\psi$&$2.0\times 10^{-15}{}^{\ *} $\\$
 -\frac{1}{2} k_{\text{eff}} (\delta v_{z1}+\delta v_{z2}) \delta \Omega_y \Omega_y T^3 \cos\theta \cos\psi$&$1.5\times 10^{-15} {}^{\ *}$\\$
 \frac{25}{24} k_{\text{eff}}(T_{\text{xx}}+t_\text{xx}) (\delta v_{z1}+\delta v_{z2}) \Omega_y^2 T^5  \cos\theta \cos\psi$&$1.5\times 10^{-15}{}^{\ *} $\\
$ -\frac{17}{12} k_{\text{eff}}  t_\text{xz}(\delta v_{z1}+\delta v_{z2}) \Omega_y T^4 \cos\theta \cos\psi$&$1.3\times 10^{-15} {}^{\ *}$\\$
 -\frac{1}{4} k_{\text{eff}}^2 t_\text{zzz} \hbar  L  T^3\cos\theta \cos\psi (2+\cos\theta \cos\psi) \left(\frac{1}{m_{85}}-\frac{1}{m_{87}}\right)$&$1.2\times 10^{-15}  {}^{\ D}$\\
$\frac{7}{12} k_{\text{eff}}^3 t_\text{zzz} \hbar ^2 T^4 \cos\theta \cos\psi  \left(1+\cos\theta\cos\psi  \right) \left(\frac{1}{m_{85}^2}-\frac{1}{m_{87}^2}\right)$&$8.8\times 10^{-16} {}^{\ E}$\\
$ -\frac{5}{6} k_{\text{eff}}  t_\text{xy} (\delta v_{y1}+\delta v_{y2}) \Omega_y T^4\cos\theta \cos\psi$&$7.6\times 10^{-16} {}^{\ *}$\\$
 \frac{1}{8} k_{\text{eff}} (T_{\text{zz}}+t_\text{zz})^2(\delta v_{z1}+\delta v_{z2}) T^5  \cos\theta \cos\psi$&$6.8\times 10^{-16} {}^{\ *}$\\$
 -k_{\text{eff}} \Omega_y (\delta \Omega_z (\delta y_1+\delta y_2)+\delta \Omega_y (\delta z_1+\delta z_2)) T^2 \cos\theta \cos\psi$&$5.9\times 10^{-16} {}^{\ G}$\\$
 -\frac{17}{12} k_{\text{eff}} (\delta v_{z1}+\delta v_{z2}) \Omega_y^4 T^5 \cos\theta \cos\psi$&$5.4\times 10^{-16} {}^{\ *}$\\$
 -\frac{1}{2} k_{\text{eff}} t_\text{xx}(\delta x_1+\delta x_2) T^2  \sin\theta$&$4.0\times 10^{-16} {}^{\ *}$\\$
 \frac{1}{2} k_{\text{eff}}t_\text{yy} (\delta y_1+\delta y_2) T^2  \cos\theta \sin\psi$&$4.0\times 10^{-16}{}^{\ *} $\\$
 -k_{\text{eff}} (\delta v_{z1}+\delta v_{z2}) \delta \Omega_y T^2 \sin\theta$&$2.6\times 10^{-16} {}^{\ *}$\\$
 \frac{49}{8} k_{\text{eff}} (\delta v_{y1}+\delta v_{y2}) \delta \Omega_x \Omega_y^2 T^4 \cos\theta \cos\psi$&$2.1\times 10^{-16}  {}^{\ H}$\\
$ \frac{1}{6} k_{\text{eff}}^3  t_\text{zzz} \hbar ^2 T^4\cos^3\theta \cos^3\psi \left(\frac{1}{m_{85}^2}-\frac{1}{m_{87}^2}\right)$&$1.3\times 10^{-16}  {}^{\ I}$\\
\end{tabular}
\caption{Phase terms with relative magnitude $>10^{-16}$ after $\vec{k}$-reversal and internal state modulation (operation of Bragg AI sequentially in two hyperfine states for suppressing magnetic field sensitivity~\cite{QTEST}), without gravity modulation. 
Parameters are defined in Table~\ref{tab3}.
Letter superscipts of the relative magnitudes indicate the corresponding phase terms in Table~\ref{tab2}, while asterisk superscripts show $<10^{-16}$ contribution of each after modulation.}\label{tab1}
\end{table*}
\begin{table*}
\begin{tabular}{c| c}
Phase term & Relative magnitude \\
\hline
\hline
$
 -4 k_{\text{eff}} (\delta v_{y1}+\delta v_{y2}) \delta \Omega_z \Omega_y T^3 \cos\theta \cos\psi $&$1.2\times10^{-14}{}^{\ A}$\\$
 \frac{1}{2} k_{\text{eff}} T_\text{yz} ((\delta y_1+\delta y_2)+(\delta v_{y1}+\delta v_{y2}) T) T^2 \cos\theta \cos\psi$&$3.3\times10^{-15}{}^{\ B}$\\$
 -\frac{1}{2} k_{\text{eff}} t_\text{zz} ((\delta v_{x1}+\delta v_{x2}) \epsilon_1+(\delta v_{y1}+\delta v_{y2}) \epsilon_2) T^3 \cos\theta \cos\psi$&$8.1\times10^{-16}{}^{\ C}$\\$
  -\frac{1}{8} k_{\text{eff}}^2 t_\text{zzz} \hbar  L  T^3\cos\theta \cos\psi (2+\cos\theta \cos\psi) \left(\frac{1}{m_{85}}-\frac{1}{m_{87}}\right)$&$5.9\times 10^{-16}{}^{\ D} $\\$
\frac{7}{12} k_{\text{eff}}^3 t_\text{zzz} \hbar ^2 T^4 \cos\theta \cos\psi \left(\frac{1}{m_{85}^2}-\frac{1}{m_{87}^2}\right)$&$4.4\times 10^{-16} {}^{\ E} $\\
$ \frac{1}{2} k_{\text{eff}} t_\text{xx}(\delta v_{x1}+\delta v_{x2}) \epsilon_1 T^3  \cos\theta \cos^2\frac{\psi }{2}$&$4.0\times10^{-16} {}^{\ F}$\\$
 -k_{\text{eff}} \delta \Omega_z \Omega_y (\delta y_1+\delta y_2) T^2 \cos\theta \cos\psi$&$3.0\times10^{-16}{}^{\ G}$\\$
 \frac{49}{8} k_{\text{eff}} (\delta v_{y1}+\delta v_{y2}) \delta \Omega_x \Omega_y^2 T^4 \cos\theta \cos\psi$&$2.1\times10^{-16} {}^{\ H}$\\$
 \frac{1}{6} k_{\text{eff}}^3t_\text{zzz}  \hbar ^2 T^4 \cos^3\theta \cos^3\psi \left(\frac{1}{m_{85}^2}-\frac{1}{m_{87}^2}\right)$&$1.3\times10^{-16} {}^{\ I}$\\

\end{tabular}
\caption{Phase terms with relative magnitude $>10^{-16}$ after $\vec{k}$-reversal and internal state modulation, with gravity modulation. 
Superscipts of the relative magnitudes indicate the corresponding phase terms in Table~\ref{tab1}. }\label{tab2}
\end{table*}

\begin{table*}
\begin{tabular}{c|c| c}
Parameter & Description & Value \\
\hline
\hline
$T$& AI interrogation time & 10~s\\
$\Omega_y$ &  ISS angular velocity & 1.13~mrad/s\\
$\delta\Omega_x,\delta\Omega_y,\delta\Omega_z$ & rotation compensation error & 1.13~$\mu$rad/s\\
$T_\text{xx},T_\text{yy},T_\text{zz}$ & Earth's gravity gradient (diagonal) & $(-0.5,-0.5,1)\times2568$~nm/s$^2$/m\\
$T_\text{xy},T_\text{xz},T_\text{yz}$& Earth's gravity gradient (off-diagonal) & 0.001~$T_\text{zz}$\\
$t_\text{xx},t_\text{yy},t_\text{zz}$ & gravity gradient of the rotating part (diagonal) & $3500$~nm/s$^2$/m\\
$t_\text{xy},t_\text{xz},t_\text{yz}$& gravity gradient of the rotating part (off-diagonal) & 0.01~$t_\text{zz}$\\
$T_\text{zzz}$ & $3^{\text{rd}}$ order derivative of Earth's potential& $-1.13\times 10^{-12}/$s$^2/$m\\
$t_\text{zzz}$ & $3^{\text{rd}}$ order derivative of the rotating potential& $-10\times 10^{-12}/$s$^2/$m\\
$\theta,\phi$& misalignment angle of $\vec{k}_\text{eff}$ to vertical in the $x$-, $y$-direction & 1~mrad\\
$\delta x_i,\delta y_i,\delta z_i$& initial position mismatch at source location $i\ (i=1,2)$& 1~$\mu$m\\
$\delta v_{xi},\delta v_{yi},\delta v_{zi}$ & initial velocity mismatch at source location $i\ (i=1,2)$& 1~$\mu$m/s\\
$x_t,y_t,z_t$& location of the center of mass of the rotating part & 0.2~m\\
$L$& separation of the two source locations & 0.5~m\\
$\epsilon_1,\epsilon_2$& angular error of turntable flipping in the $x$-, $y$-direction & 0.1~mrad$^{*}$\\
\end{tabular}
\caption{Definition of parameters in the simulation. 
The Earth's gravity potential is expanded at 400~km altitude as: $-\left(g z+\frac{1}{2}T_{\text{zz}}z^2+\frac{1}{3!}T_{\text{zzz}}z^3+\frac{1}{2}T_{\text{xx}}x^2+\frac{1}{2}T_{\text{yy}}y^2+T_{\text{xy}}xy+\cdots\right)$.
The gravitational potential of the rotating part is modeled as: $-\left(\frac{1}{2}t_{\text{zz}}(z-z_t)^2+\frac{1}{3!}t_\text{zzz}(z-z_t)^3+t_{\text{xy}}(x-x_t)(y-y_t)+\cdots\right)$.
When the turntable points in the $+z$-direction, source location 1 is near $(0,0,0)$ and source location 2 is near $(0,0,L)$.
A stationary dual species cloud at each source location is first launched with momentum $\pm\hbar\vec{k}_\text{eff}$ toward the other source location; after $T/2$ the interferometer pulse sequence is then applied with pulse separation time $T$.
Flipping of the turntable is about the $y$-axis at $(0,0,L_t)$ with $L_t=0.3$~m.
Off-diagonal gradient elements are employed to model 1~mrad misalignment of the instrument to the principal axes of $\overset{\text{\tiny$\bm\leftrightarrow$}}{\gamma}_\oplus$ and 10~mrad to the principal axes of $\overset{\text{\tiny$\bm\leftrightarrow$}}{\gamma}_{\textrm{S}}$, respectively.
Note that the contribution of shot-to-shot fluctuations of initial conditions, e.g. due to the cloud position and velocity profiles, is below atom shot noise as detailed in Ref.~\cite{QTEST}.
$^{*}$Space-qualifiable turntables with $<10\mu$rad repeatability and accuracy are available, e.g., the Aerotech ALAR series~\cite{turntable}.
}\label{tab3}
\end{table*}

To further substantiate our arguments, a simulation of phase shifts and cancellations is performed in the setting of QTEST~\cite{QTEST}, a concept of an apparatus on a turntable aboard ISS running simultaneous $^{85}$Rb and $^{87}$Rb AIs (each with 500:1 signal-to-noise ratio per shot governed by $10^6$ atoms and 50\% contrast) for an EP test, shown schematically in Fig.~\ref{fig1}.
QTEST is designed to achieve $\eta \le 10^{-15}$ in 1 year of  continuous operation, thanks to the insensitivity to vibrations and thus not interfering with ISS operations or astronaut activities.
In QTEST, there is a dual species source at each end of a magnetically shielded science chamber, where the sources and the chamber are mounted on a turntable orienting the chamber toward or away from the Earth.
At each turntable orientation, dual AIs are launched from each source toward the other end, driven by common Bragg pulses retroreflected along the science chamber to facilitate simultaneous $\vec{k}$-reversal measurements.
The experimental sequence is operated at 70 s duty cycle, and the orientation of the turntable is changed every 10 runs~\cite{QTEST}.
The simulation is conducted in the ISS frame, and four scenarios are included to account for dual source regions and two turntable positions.
In each scenario, the phase difference $\Delta \Phi$ between $^{85}$Rb and $^{87}$Rb is calculated and expressed in terms of initial conditions, parameters of gravitational and magnetic potentials, and pointing change of $\vec{k}_{\rm{eff}}$ for rotation compensation~\cite{QTEST,Hogan2008light}.
A combination of phase differences features both $\vec{k}$-reversal and turntable modulation cancellation.

The simulation result is summarized in Tables~\ref{tab1} and~\ref{tab2}, where relative magnitude is defined as the size of the corresponding phase term over $k_\text{eff} g T^2$ for $T=10$~s.
Table~\ref{tab1} lists residual phase terms up to $10^{-16}$ relative magnitude after $\vec{k}$-reversal and internal state modulation~\cite{QTEST}, without gravity modulation, and Table~\ref{tab2} lists residuals after additional gravity modulation with flipping angle errors $\epsilon_1,\epsilon_2$. 
The error terms induced by SGG are all proportional to the flipping error and $<10^{-15}$ with the simulation parameters listed in Table~\ref{tab3}, where the relevance and justification of the parameter values are discussed in detail in~\cite{QTEST}.
The largest two terms in Table~\ref{tab2}, both of which are aligned to the turning axis and not modulated, can be suppressed with the imaging technique outlined in Ref.~\cite{QTEST} by at least $10^2$, and thus will not limit the performance of QTEST at $10^{-15}$.

The above simulation assumes that the initial condition mismatches are 100\% fixed to the apparatus.
Practical imperfections, e.g., assuming that 0.01\% of $\delta x$ and $\delta v$ are not fixed to the apparatus due to gravitational and magnetic fields outside the turning apparatus, will effectively leave the same proportion in amplitude of phase terms in Table~\ref{tab1} unmodulated and not suppressed.
This is equivalent to additional errors of $10^{-4}$ of Table~\ref{tab1}, resulting in additional error contribution $<10^{-15}$.
Since $^{85}$Rb and $^{87}$Rb have the same linear Zeeman shift
, the magnetic field bias itself doesn't contribute to displacement.
The differential gravitational sag $\delta z_{\textrm{sag}}$, which is the weight difference over the spring constant $m \omega^2$ in the proposed quadrupole-Ioffee configuration (QUIC) trap~\cite{QTEST} for QTEST, is:
\begin{equation}
\delta z_{\textrm{sag}}=\frac{\delta m}{m} \frac{ a_{\rm{res}}}{\omega^2},
\end{equation}where $\delta m/m\simeq 0.02$ for Rb isotopes, $a_{\rm{res}}$ is the residual acceleration in microgravity, and $\omega\simeq 2\pi\times 25$ Hz will be the trap frequency.
To constrain $\delta z_{\textrm{sag}}\le0.1$ nm due to external gravitational force, the gravity at the AI needs to be $a_{\rm{res}}\le 0.12$ mm/s$^2$, corresponding to an altitude difference of the apparatus to the center of mass of the ISS of $a_{\rm{res}}/(\Omega_{\textrm{ISS}}^2+\gamma_\oplus)\le 31.7$ m, which is easily satisfied.

The fundamental limit of the position-velocity uncertainty of individual atoms, governed by the Heisenberg's uncertainty principle, is orders of magnitude smaller ($10^{-3}$ for $10^6$ atoms) than the atom-shot-noise-limited uncertainty of the ensemble centroid in each measurement~\cite{Nobili}. 
The uncertainty in determining the centroid of a source, governed by the central limit theorem, decreases as the number of measurements increases.
This statistical nature of both uncertainty limits suggests that the atom shot noise will always dominate for the thermal clouds considered in this letter. 

Our result is consistent with the gradient effect suppression scheme adopted in the \mbox{MICROSCOPE} mission where bulk test masses are used with a target sensitivity of $10^{-15}$~\cite{uscope,touboul2012microscope}, despite that \mbox{MICROSCOPE} has a fixed and characterized overlap mismatch while QTEST has random but specified mismatch tolerances. 
The EP signal in \mbox{MICROSCOPE} is modulated at the orbiting frequency (or the orbiting plus the spinning frequencies in the spinning mode), while the gravity gradient signal is mostly at twice the frequency.
We interpret the result in this letter a direct consequence of the fact that gravity gradients don't change sign under flipping as shown in Eq.~(\ref{eq6}).

In summary, we report that the gravity gradient dependent systematics, including those due to the self gravity gradient, can be totally suppressed in AI-based EP tests when the apparatus is inverted by a gimbal.
This suppression is based on the fixation of initial condition mismatch of two species to the apparatus under microgravity, and on the fact that gravity gradient is the second order derivative of a scalar function so that the sign remains the same when inverted.
We discuss the physics in both the Earth frame and the apparatus frame reaching the same conclusion, which is supported by a more elaborated simulation.
We conclude that, thanks to this suppression scheme, QTEST can reach the targeted EP sensitivity of $10^{-15}$.

We thank Brian Estey for his contributions on theory verification. This work was carried out at the Jet Propulsion Laboratory, California Institute of Technology, under a contract with the National Aeronautics and Space Administration. Government sponsorship acknowledged.

\bibliographystyle{unsrt}
\bibliography{QTESTGG}

\end{document}